# System Evaluation of the Intel Optane Byte-addressable NVM

Ivy B. Peng
peng8@llnl.gov
Lawrence Livermore National Laboratory
Livermore, USA

Maya B. Gokhale
gokhale2@llnl.gov
Lawrence Livermore National Laboratory
Livermore, USA

Eric W. Green
green77@llnl.gov
Lawrence Livermore National Laboratory
Livermore, USA

## ABSTRACT

Byte-addressable non-volatile memory (NVM) features high density, DRAM comparable performance, and persistence. These characteristics position NVM as a promising new tier in the memory hierarchy. Nevertheless, NVM has asymmetric read and write performance, and considerably higher write energy than DRAM. Our work provides an in-depth evaluation of the first commercially available byte-addressable NVM – the Intel Optane® DC™ persistent memory. The first part of our study quantifies the latency, bandwidth, power efficiency, and energy consumption under eight memory configurations. We also evaluate the real impact on in-memory graph processing workloads. Our results show that augmenting NVM with DRAM is essential, and the combination can effectively bridge the performance gap and provide reasonable performance with higher capacity. We also identify NUMA-related performance characteristics for accesses to memory on a remote socket. In the second part, we employ two fine-grained allocation policies to control traffic distribution between DRAM and NVM. Our results show that bandwidth spilling between DRAM and NVM could provide 2.0x bandwidth and enable 20% larger problems than using DRAM as a cache. Also, write isolation between DRAM and NVM could save up to 3.9x energy and improves bandwidth by 3.1x compared to DRAM-cached NVM. We establish a roofline model to explore power and energy efficiency at various distributions of read-only traffic. Our results show that NVM requires 1.8x lower power than DRAM for data-intensive workloads. Overall, applications can significantly optimize performance and power efficiency by adapting traffic distribution to NVM and DRAM through memory configurations and fine-grained policies to fully exploit the new memory device.

**ACM Reference Format:**
Ivy B. Peng, Maya B. Gokhale, and Eric W. Green. 2019. System Evaluation of the Intel Optane Byte-addressable NVM. In *Proceedings of* . ACM, New York, NY, USA, 12 pages.

## 1 INTRODUCTION

A diversity of applications on HPC and cloud computing systems demand ever-increasing memory capacity to enable expanding workloads. In recent years, HPC applications have been observed to converge towards "Big Data" because of the enormous amount of data sets [27]. Neural networks in machine learning applications can improve accuracy by using wide and deep networks [29, 31], but network complexity may be restricted by the memory capacity of a single machine. Large-scale graphs often have to be distributed over multiple compute nodes to enable in-memory processing [19]. Simply scaling up the memory capacity using the DRAM technology can be prohibitively expensive in both power and cost. As a volatile memory technology, DRAM requires power to refresh data periodically, and the refresh power scales proportionally with the memory capacity. In fact, the power constraint has been identified as one of the main challenges in Exascale computing [2]. Moreover, DRAM faces challenges in further scaling down the size of capacitors and transistors, and the low density makes it infeasible for implementing large-capacity systems within area constraints [17].

Non-volatile memory (NVM) technologies are considered as a promising alternative to DRAM for its high density, low standby power, and low cost per bit. Nevertheless, their access latency could be as high as 3 – 20 times that of DRAM. Additionally, their low bandwidth, asymmetric read and write performance, and high write energy hinder their suitability as the primary system main memory. Recently, a byte-addressable NVM using the Intel Optane® DC™ technology (shortened to Intel Optane DC PMM) has become commercially available, enabling up to 6 TB capacity on a single machine [13]. While previous works have studied NVM technologies using simulations and emulations [5, 14, 17, 25, 26], a realistic evaluation on the hardware enables accurate assessment of its impact on applications and future system designs. In this work, we perform extensive experiments and modeling to identify the main consideration for adapting applications for utilizing the new memory device efficiently.

Our study consists of two main parts. First, we quantify the performance, power, and energy consumption under eight memory configurations that require no application modifications. We choose five graph applications from GAP [1] and Ligra [28] framework to evaluate the efficiency of memory configurations. Our results show that using DRAM as a cache to NVM can effectively bridge the performance gap and brings performance close to DRAM. We also show that using local NVM on a single socket may be more efficient than using DRAM on two sockets for some workloads. However, directly replacing DRAM with NVM for graph applications could decrease performance by an order larger than the gap between DRAM and NVM in bandwidth and latency.

The second part of our study employs a set of allocation policies to enable fine-grained traffic distribution between DRAM and NVM. In particular, we highlight the importance of bandwidth spilling between DRAM and NVM. We also quantify the performance improvement and energy saving by isolating write-intensive data structures to DRAM compared to DRAM-cached NVM. Finally, we establish the roofline model of the theoretical peak performance [4, 33] to explore power efficiency at different traffic distribution. Our results show that the Intel Optane DC PMM can improve power and energy efficiency when traffic distribution is adapted to various arithmetic intensities. Our main contributions are as follows:

- Quantify the latency, bandwidth, power, and energy consumption of eight memory configurations for diverse access patterns
- Identify the impact on bandwidth and power efficiency from non-temporal writes in DRAM-cached NVM
- Evaluate the efficiency of using DRAM as a cache to NVM for large-scale graph workloads
- Identify the advantage of using the Optane PMM on the local socket to avoid performance loss from accessing DRAM on a remote socket
- Propose a DRAM-NVM bandwidth spilling allocation policy to achieve 2.0x bandwidth and enable larger problems than DRAM-cached NVM.
- Quantify that a write-isolation policy between DRAM and NVM can save up to 3.9x energy and improves bandwidth by 3.1x compared to DRAM-cached NVM.
- Establish the roofline model of theoretical peak performance at various distributions of read-only traffic between DRAM and NVM and show that a balanced distribution could improve performance and power efficiency.

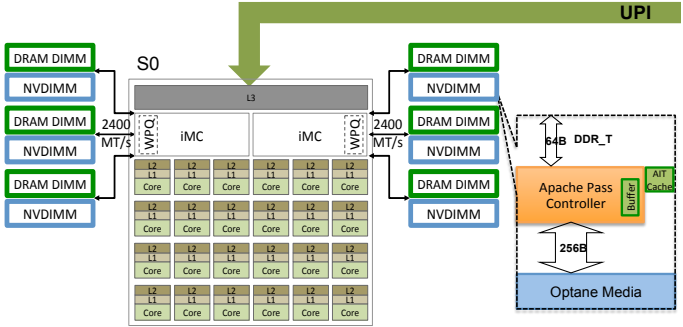

Figure 1: A conceptual diagram of one socket on the Purley Platform. Each socket consists of two memory controllers and six channels attached to DRAM DIMMs and NVDIMMs. Data in the write pending queue (WPQ) in iMC will be flushed to NVDIMMs even during power failure. NVDIMM includes small DRAM caches (green boxes) for caching data.

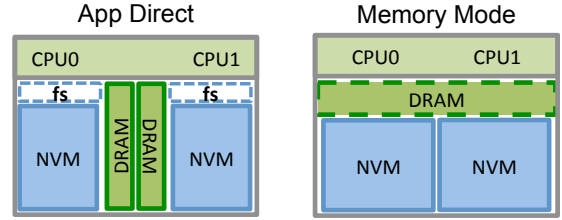

Figure 2: The logical view of configuring all NVDIMMs either in App Direct mode or Memory mode.

## 2 ARCHITECTURE

Our evaluation of the Intel Optane byte-addressable NVM uses the Purley platform. The platform consists of two sockets that feature two 2$^{nd}$ Gen Intel® Xeon® Scalable processors and Intel Optane DC PMM[1]. Figure 1 presents an overview of the socket architecture. Each socket has two integrated memory controllers (iMC) that control six memory channels, which are attached to DRAM DIMMs and NVDIMMs. An NVDIMM can have a capacity of 128, 256 or 512 GB. Currently, 128-GB NVDIMM has the lowest cost per byte [7].

NVDIMMs use a non-standard DDR-T protocol to enable out-of-order commands and data transfer to address the long latency to Optane media [7]. In contrast, DRAM DIMMs employ the standard DDR4 protocol. The inset on Figure 1 illustrates the different data granularity between CPU and NVDIMM (64 bytes) and the Optane media (256 bytes). A small DRAM buffer is used to cache data from the media so that consecutive requests to the same 256 byte could have reduced latency. There is a small processor (Apache Pass controller) in NVDIMM to translate physical addresses into internal Optane device addresses [13] and caches the address indirection table (AIT) in a DRAM cache for performance.

A common configuration of the memory system attaches a DRAM DIMM and an NVDIMM to one memory channel. Having an iMC connected to both DRAM DIMMs and NVDIMMs is essential for enabling DRAM caching to NVM because DRAM can only cache accesses to NVDIMMs that share the same iMC [13]. Under this channel-sharing configuration, the aggregate bandwidth from DRAM and NVM becomes unachievable. In comparison, the Intel Knights Landing processor has separate memory channels for DRAM and MCDRAM, and peak bandwidth could be an aggregate of both memories.

**Memory Options** Intel Optane DC PMM can be either configured in *App Direct* or *Memory* mode. Through the *ipmctl* utility [10], users can select the configuration for each NVDIMM so that the platform could be configured either in one mode or a hybrid mode. Figure 2 presents the logical view when all NVDIMMs are configured in the same mode. In App Direct mode, DRAM on the two sockets are exposed as a shared memory with two non-uniform memory access (NUMA) nodes. Separate namespaces [11] are created for PMM on the two sockets using *ndctl* utility [12].[2] In Memory mode, the Optane memory modules on the two sockets are visible as two NUMA nodes to CPUs while DRAM becomes a transparently managed cache.

In App Direct mode, a *dax*-aware file system would transparently convert file read and write operations into 64-byte load and store instructions. Although the interaction between the host processor and NVDIMM is now at a much finer granularity than in block devices, each request still fetches 256 bytes from the media. Thus, data locality that utilizes all the fetched bytes would bring optimal performance. Similarly, writes to PMM are performed in 256 bytes.

---

[1]We use PMM and NVM interchangeably for the rest of the paper.

[2]Ndctl fails to create one namespace for all NVDIMMs on two sockets likely because their memory addresses are non-contiguous.

System Evaluation of the Intel Optane Byte-addressable NVM

Modifying fewer than 256 bytes still incurs the overhead and energy as of 256 bytes, i.e., *write amplification*.

Large PMM capacity also results in large metadata for the page table. Per-page metadata occupies 64 bytes when creating namespaces for PMM. Even using the smallest 128-GB NVDIMMs on the platform would result in 24 GB metadata. Users can choose to store this metadata in DRAM or PMM. However, we find that storing metadata in PMM could severely impact performance.

In Memory mode, DRAM becomes a direct-mapped write-back cache to PMM and can only cache accesses to NVDIMMs attached to the same iMC [13]. One impact of this cache mechanism is that DRAM on one socket cannot cache accesses to PMM on the other socket, which contributes to NUMA effects in Memory mode [6]. This design is likely a trade-off between flexibility and performance to avoid routing requests among iMCs. As a write-back cache, writes are automatically buffered in DRAM, which is critical for avoiding performance degradation due to low write bandwidth to PMM. The platform provides two options in the optimization mode, i.e., for latency or for bandwidth, as a BIOS setting. We find that the option shows impact in Memory mode for large data size.

**Memory Power** DRAM and NVM are tightly coupled on the system for performance and could impact the power consumption of the system. Each NVDIMM includes small DRAM caches, and a controller for address translation and write-leveling management [13]. These components consume additional active and static power, even though the non-volatile media does not require standby power to refresh data. At the system level, the metadata of namespaces needs to be stored in DRAM instead of PMM to avoid significant slowdown. Also, Memory mode relies on DRAM DIMMs to cache accesses to PMM. Therefore, (at least some) DRAM DIMMs need to be power-on for acceptable performance of PMM. In this study, we evaluate the power and energy consumption of Optane PMM for realistic workloads under different memory configurations.

**Application Porting** Utilizing PMM in App Direct mode requires porting efforts to select data structures in applications and change their allocation sites. A variant of the App Direct mode is to expose PMM to the kernel as separate NUMA nodes. In this configuration, standard NUMA control techniques like *numactl* utility can enable applications utilizing PMM without any modifications. Also, using DRAM as a cache in Memory mode requires no application changes. We believe these memory configurations that require minimal porting efforts are likely to be the initial deployment and thus, we perform an in-depth evaluation of these configurations to provide insights for selecting the optimal configuration for a workload, and also for avoiding combinations of access patterns and configurations that could cause severe performance bottlenecks.

## 3 METHODOLOGY

In this section, we describe the experimental setup, benchmarks, applications, and methodologies. Table 1 specifies the configuration of our testbed. We always configure 12 NVDIMMs in the same mode. The speed of the data bus is 2400 GT/s, supporting a peak bandwidth of 19.2 GB/s per channel or 230.4 per platform.[3] Overall, the system has 192 GB DRAM and 1.5 TB NVM. We store the page metadata for the Optane PMM namespaces in DRAM, leaving 168 GB DRAM available to applications. We use a set of memory configurations as specified in Table 2 for evaluation.

---
[3]Note that higher speed could not be enabled on the platform even though DDR4 supports 2666 GT/s.

**Table 1: Experiment Platform Specifications**

| Model | Intel® Xeon® Platinum 8260L |
|---|---|
| Processor | 2$^{nd}$ Gen Intel® Xeon® Scalable processor |
| Cores | 24 Cores (48 hardware threads) × 2 sockets |
| Speed | 2.4 GHz, 3.9 GHz Turbo frequency |
| L1 Cache | 32 KB d-cache and 32 KB i-cache (private) |
| L2 Cache | 1 MB (private) |
| L3 Cache | 35.75 MB (shared) |
| TDP | 165 W |
| Memory Controller | 2 iMCs × 3 channels × 2 sockets |
| DRAM | 16-GB DDR4 DIMM per channel |
| NVM | 128-GB Optane DC NVDIMM per channel |
| UPI Links | three links at 10.4 GT/s, 10.4GT/s, and 9.6 GT/s |

**Table 2: Memory Configurations**

| Configuration | Optane Mode | Mapping/Namespace | Socket | Data Binding |
|---|---|---|---|---|
| DRAM-local | App Direct | memmap | local | DRAM |
| DRAM-remote | App Direct | memmap | remote | DRAM |
| PMM-numa-local | App Direct | memmap | local | PMM |
| PMM-numa-remote | App Direct | memmap | remote | PMM |
| PMM-fsdax-local | App Direct | fsdax | local | PMM |
| PMM-fsdax-remote | App Direct | fsdax | remote | PMM |
| MemoryMode-local | Memory Mode | — | local | — |
| MemoryMode-remote | Memory Mode | — | remote | — |
| DRAM | App Direct | memmap | two sockets | DRAM |
| PMM | App Direct | memmap | two sockets | PMM |
| DRAM-PMM-interleave | App Direct | memmap | two sockets | interleave all |
| MemoryMode | Memory Mode | — | two sockets | — |

The platform runs operating system Fedora 29 with GNU/Linux 5.1.0. We compile all applications using GCC 8.3.1 compiler with support for OpenMP. We use the Intel Memory Latency Checker (MLC) [9] to quantify the latency and bandwidth for benchmarking. In addition, we use the STREAM [20] benchmark and extended it to include an accumulation kernel, for quantifying memory bandwidth. The accumulation kernel is a read-only workload that sums up all elements in an array. We develop a set of benchmarks to establish roofline, power-line, and arch-line models [4, 33] for performance and energy efficiency at different traffic distribution between DRAM and NVM. We use the Intel Processor Counter Monitor (PCM) [32] to collect power and energy consumption of memory and CPU on each socket.

Our experiments use GAP [1] and Ligra [28] graph processing frameworks for evaluating the real impact of PMM on applications. We select breadth-first search (BFS), betweenness centrality (BC), triangle counting (TC), connected component (CC), and PageRank (PR) applications from each framework. The experiments use graphs generated by the included Kronecker [18] generator in GAP and the rMat [3] generator in Ligra. The largest input in Ligra (s30) has 1073M vertices and 17179M edges and requires about 625 GB memory. The largest input in GAP has 2147M vertices and 34359M edges. It requires about 1049 GB memory for TC and 540 GB memory for the other applications. If not specified, we run all applications using 96 threads on two sockets. For single-socket execution, we use 48 threads and memory on one socket, i.e., 96 GB DRAM and 768 GB NVM, to eliminate the influence from NUMA effects.

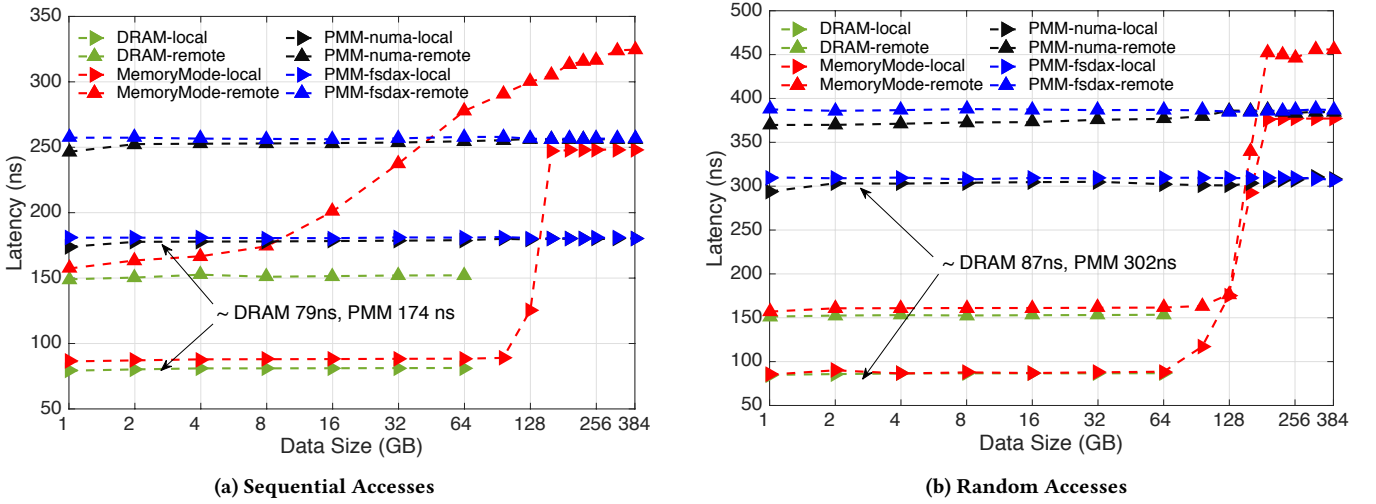

Figure 3: Latency of sequential and random read accesses to a data buffer with increasing size. DRAM capacity on a single socket is 96 GB and the total DRAM capacity of the system is 192 GB.

## 4 MEMORY CONFIGURATIONS

In this section, we focus on memory configurations that require no application modifications. We first benchmark the latency, bandwidth, and power efficiency of all configurations, and then quantify the impact on five graph applications.

### 4.1 Latency

We quantify the read latency in eight memory configurations and present the results in Figure 3. DRAM-local and PMM-numa-local configurations are two "bare-metal" configurations that are not subject to cache overhead in Memory mode or file system overhead in PMM-fsdax. Based on them, we quantify the sequential read accesses to DRAM has a latency of 79 ns and 174 ns to PMM. For random accesses, the latency to PMM increases to 302 ns while to DRAM it slightly increases to 87ns. PMM is more sensitive to data locality because the internal data granularity is at 256 bytes, and data is buffered in NVDIMM (Section 2).

MemoryMode-local has latency close to DRAM-local when the data size fits in one socket (96 GB) for both sequential and random accesses. Interestingly, for both access patterns, once the data size exceeds a single socket, the latency approaches that of PMM-numa-remote configuration. In Figure 3b, lines for DRAM-local and MemoryMode-local are nearly overlapping for data size up to 64 GB, indicating that managing DRAM as a cache incurs little overhead. MemoryMode-remote has increasing latency for sequential accesses even at small data size, likely because the local DRAM cannot cache accesses to PMM on another socket.

The *dax*-aware file system imposes very minimal overhead compared to accessing PMM as a NUMA node. PMM-numa-local and PMM-fsdax-local configurations have nearly identical latency at all data sizes in Figure 3a and 3b. Note that *dax*-aware file system implicitly converts file reads and writes into load and store instructions and bypasses the page cache in the kernel. Also, PMM-fsdax configurations can provide data persistence in case of DRAM power-off because data that has reached iMC will be flushed into PMM within DRAM retention time. Overall, the Intel Optane provides persistence at fine grain and low overhead.

NUMA effects across the two sockets have a severe impact on all memory configurations. We divide the eight configurations into four groups of local and remote configurations and present them in the same color in Figure 3. For both access patterns, NUMA effects increase latency by 1.2 to 1.8 times. We notice that the increased latency remains nearly constant for each group, in the range of 66-85 ns. Surprisingly, for sequential accesses, when the data size is as small as 16 GB, latency in Memory-remote configuration is already higher than in PMM-local configuration. Also, starting from 160 GB, MemoryMode-local has higher latency than PMM-numa-local. The high latency in MemoryMode indicates that accessing local PMM could be an alternative. When data placement control is feasible, explicitly managing data in App Direct mode to utilize local PMM may have lower latency than Memory mode.

*Insight I: Coordinating 256B accesses to PMM to exploit locality (i.e. using the PMM internal granularity) may reduce latency and write-amplification.*

*Insight II: Explicit data placement that utilizes local PMM could mitigate high cost of accessing DRAM on the remote socket.*

### 4.2 Bandwidth

We quantify the peak bandwidth of six access patterns on a single socket by scaling the number of threads. Note that increasing the number of threads beyond 24 (one thread per core) brings minimal changes to the bandwidth, and thus is not presented. We derive the bandwidth to PMM and DRAM from PMM-numa-local and DRAM-local configurations, respectively. For sequential read accesses in

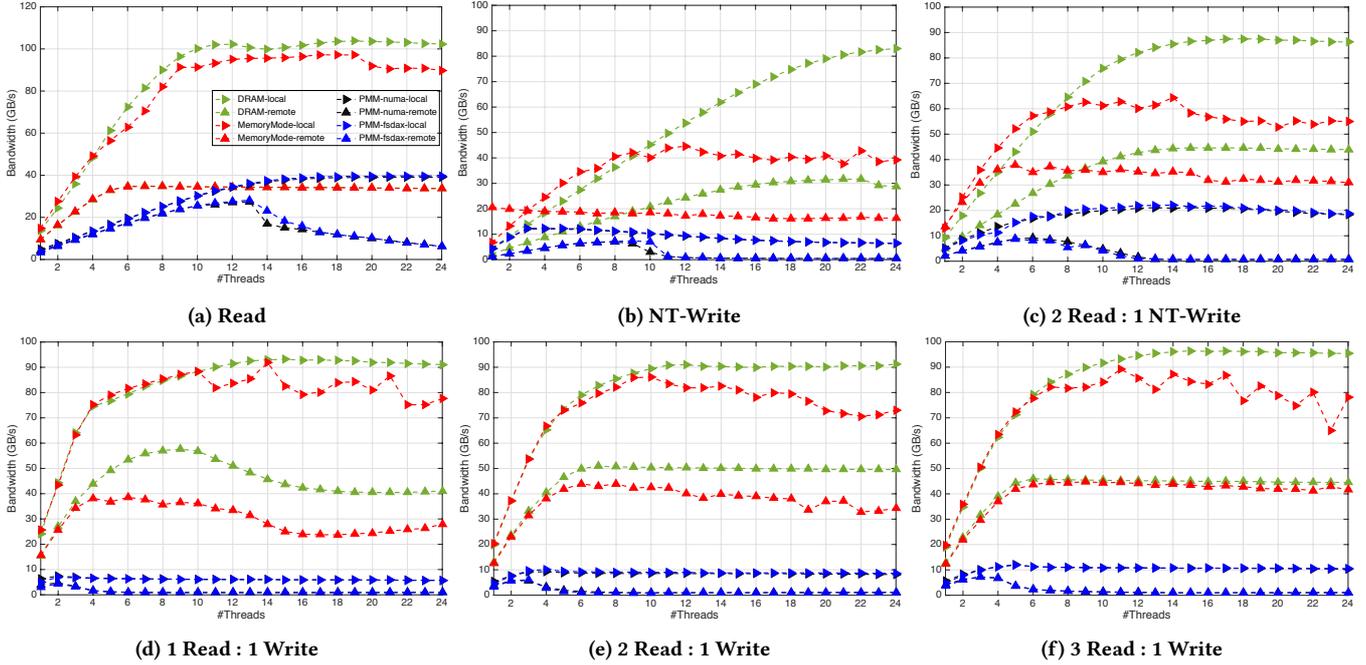

**Figure 4: Memory bandwidth of mixed read and write sequential accesses on a single socket. One thread per core is used.**

Figure 4a, DRAM has a peak bandwidth of 104 GB/s while PMM reaches 39 GB/s. Also, PMM has a 3.3 times asymmetry in read and write bandwidth, given its write bandwidth at 12.1 GB/s. The dax-aware file system shows low overhead so that PMM-fsdax and PMM-numa configurations are always overlapping in Figure 4.

Using local PMM achieves higher bandwidth than the remote DRAM for read-only traffic. PMM-numa-local and PMM-fsdax-local (overlapping black and blue lines in Figure 4a) start outperforming DRAM-remote and MemoryMode-remote (overlapping green and red lines in Figure 4a) when more than 14 threads are used. However, the performance of PMM degrades once write accesses are mixed in. In Figure 4d to 4f, the gap between DRAM and PMM increases to 4.1–12.5 times, in contrast to the 2.6 times gap in read-only accesses in Figure 4a. In these patterns, the local DRAM could still sustain 84.9–98.7 GB/s bandwidth while the bandwidth of PMM-numa-local and PMM-fsdax-local dramatically decreases to 7.6–21.6 GB/s. Interestingly, the lowest bandwidth is obtained with mixed read and write accesses rather than write-only accesses. In Figure 4d, 4e, and 4f, the bandwidth of PMM local configurations steadily increases when the ratio of read accesses increases.

Non-temporal stores [15] (NT-write) could significantly diminish the performance of Memory mode at a large number of threads. In Figure 4b and 4c, the bandwidth of MemoryMode-local is only 47% and 64% that of DRAM-local at 24 threads. Without NT-write, Memory mode could sustain 80 to 88% DRAM bandwidth in Figure 4a, 4d, 4e and 4f. Typically, non-temporal stores are used in applications to avoid caching data that will not be reused shortly to improve cache utilization. However, for Intel Optane PMM, caching writes in DRAM becomes more critical for performance. Interestingly, for a small number of threads, i.e., 8 and 9 threads in Figure 4b and 4c, MemoryMode-local with NT-write accesses outperforms DRAM-local configuration.

NUMA access further exacerbates the bandwidth to PMM, causing severe performance degradation in PMM-numa-remote and PMM-fsdax-remote configurations. In Figure 4d to 4f, when more than three threads are used, the bandwidth to the remote PMM starts decreasing, eventually reaching below 1GB/s. Although the links between the two sockets have a high aggregated bandwidth, the measured bandwidth is far below the peak, implying significant contention on the links. Mitigating such performance loss becomes a priority. Intelligent co-location of data and computation on the same socket and utilizing local PMM is more effective than reaching over the link. Moreover, throttling concurrent remote accesses could also mitigate performance degradation.

MemoryMode-local configuration exhibits reduced bandwidth and increased variation in performance as the number of threads increases. In Figure 4, the gap between DRAM-local and MemoryMode-local continues increasing when more than 10 threads are used. Note that the total data size in these tests is smaller than the DRAM capacity on a single socket. Therefore, the increased bandwidth loss is likely due to the increased cache conflicts in DRAM. Since DRAM is configured as a direct-mapped cache, when multiple threads concurrently access DRAM, the probability that multiple threads fetch different data that is mapped into the same cache set also increases. Consequently, for such problem sizes, DRAM-local would be more suitable than MemoryMode-local configuration.

MemoryMode-local configuration is also highly sensitive to the optimization mode for bandwidth or latency (Section 2). Figure 5

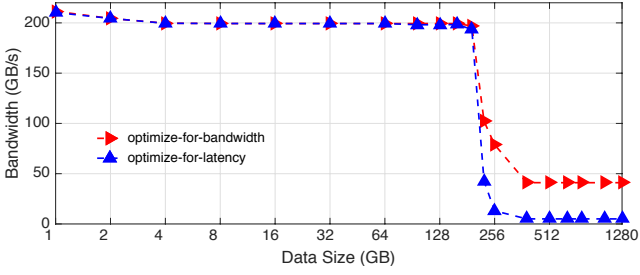

Figure 5: Memory bandwidth on two sockets in Memory-Mode using the optimization mode for bandwidth and latency respectively.

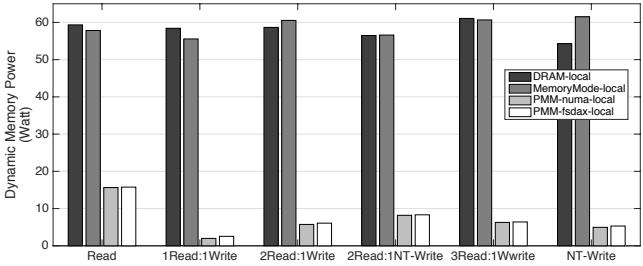

Figure 6: Dynamic memory power of a single socket for six read write mixed workloads under four local configurations.

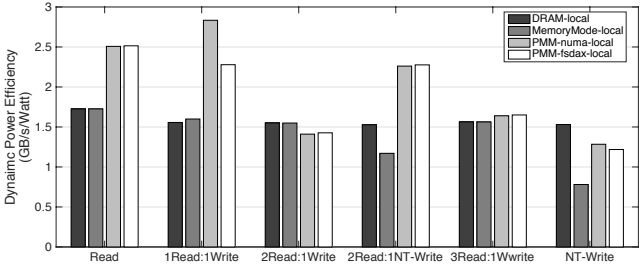

Figure 7: Dynamic memory power efficiency calculated as bandwidth per dynamic memory power.

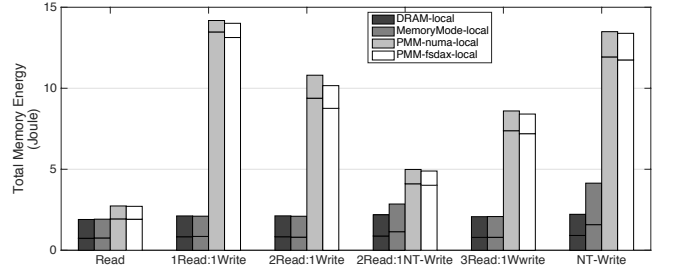

Figure 8: Total memory energy with breakdown into static (the bottom partition) and dynamic energy (the top partition). Note that the static energy consists of DRAM static energy because DRAM DIMMs cannot be switched off.

presents the sequential read bandwidth in MemoryMode-local configuration as the data size increases. The gap between the two options starts appearing when the data size exceeds the total DRAM capacity (192GB). Eventually, at large data size, the option for bandwidth saturates at 40 GB/s while the option for latency sustains at only 5 GB/s.

*Insight III: Local PMM could have higher bandwidth than remote DRAM for read-intensive workloads at high thread counts.*

*Insight IV: Throttling concurrent updates to PMM and isolating write-intensive data structures from PMM could optimize performance.*

### 4.3 Power and Energy
In this section, we quantify the dynamic power and power efficiency of PMM under six access patterns. We use PCM to measure the total memory power and energy consumption for each socket. One challenge is to isolate static power from the measured total power. Since each memory channel is attached with one DRAM DIMM and one NVDIMM, the measured power always includes DRAM static power because DRAM needs to refresh data periodically even without any data accesses. Our solution is to bind application execution to one socket and measure the memory power of the busy socket and idle socket, respectively. We find that the idle socket consumes nearly constant memory power of 38 Watt. Note that without running any applications, a socket consumes about 18-20 Watt memory power. The additional 18-20 Watt is likely due to activities for supporting cache lookup and coherence. Therefore, we find the idle socket power 38 Watt as a more reasonable reference to the static power at run time. Next, we subtract this static memory power from the total memory power of the busy socket to quantify dynamic memory power.

PMM significantly reduces the dynamic memory power compared to DRAM in all tested access patterns. Figure 6 presents the dynamic memory power on one socket. The PMM-numa-local and PMM-fsdax-local configurations consume similar power across all workloads. In general, the power consumed by the PMM configurations closely follows the changes in bandwidth. For instance, from 1 read : 1 write to 3 read : 1 write, the bandwidth in PMM configurations steadily increases in Figure 4, and so does the power in Figure 6, which increases from 2 to 8 Watt. In contrast, DRAM-local and MemoryMode-local configurations exhibit little change in dynamic memory power across the access patterns, stabilizing at about 60 Watt. Overall, PMM configurations reduce dynamic power by 4–29 times compared to DRAM configurations.

PMM also achieves higher or comparable power efficiency compared to DRAM in all tested workloads, except the write-only workload. We define the power efficiency as the peak bandwidth at 24 threads per socket (one thread per core) per dynamic memory power and report in Figure 7. For the read-only workload in Figure 7, PMM-numa-local and PMM-fsdax-local configurations achieve up to 47% higher power efficiency than DRAM-local. As expected, due to the high write energy to PMM, the power efficiency of PMM configurations is 20% lower than DRAM-local configuration for the write-only workload. This observation restates the importance of isolating writes from PMM and also shows the potential of using PMM for meeting a low power envelope on large-scale systems.

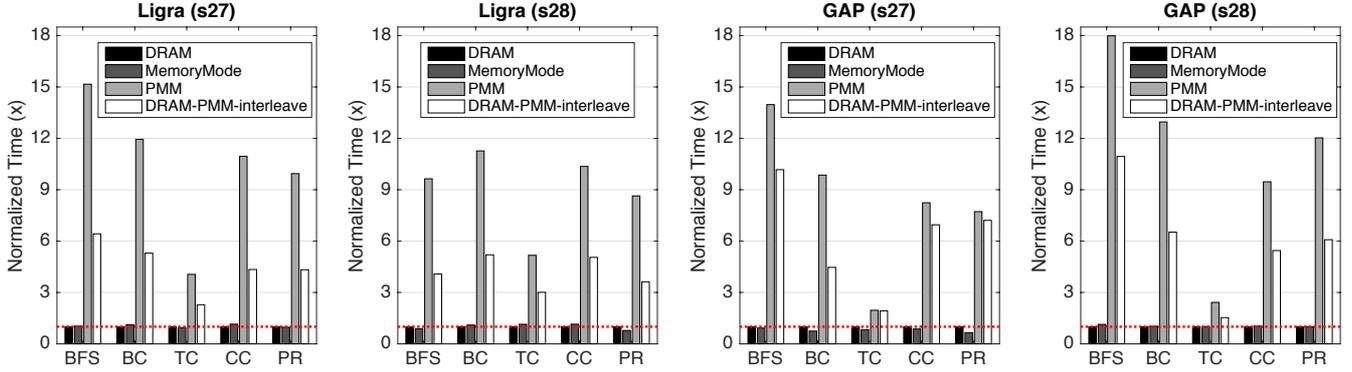

Figure 9: Normalized execution time of five graph applications (x-axis) in Ligra and GAP frameworks using four memory configurations. Time is normalized to the DRAM configuration (the red dotted line).

Non-temporal writes again significantly impact Memory mode. Without NT-write, the MemoryMode-local configuration consumes similar dynamic power as the DRAM-local configuration. However, with NT-write accesses, MemoryMode-local consumes 13% additional dynamic power. This impact is even more profound in power efficiency. The MemoryMode-local configuration shows 49% lower power efficiency than the DRAM-local configuration for NT-write accesses while it can reach similar efficiency for all other access patterns. In fact, the power efficiency in the MemoryMode-local configuration with NT-write accesses is even lower than directly writing to PMM. Consequently, the MemoryMode-local configuration consumes more memory energy than the DRAM-local configuration, as reported in Figure 8. This finding is consistent with the conclusion in the bandwidth evaluation to avoid non-temporal writes when DRAM is configured as a cache to PMM.

PMM configurations can reduce the dynamic memory energy for certain workloads. However, the high static power, which is partially because DRAM DIMMs cannot be powered off, results in high total energy costs. For bandwidth-bound workloads, the PMM configurations require longer execution time than the DRAM configurations. Despite the low dynamic power, the static power persists, and the static energy becomes dominant. Figure 8 presents the breakdown of total memory energy. The 1 read : 1 write workload spends 95% Joule for static energy. For most access patterns, the dynamic memory energy (the top partition) only takes up a small portion of the total energy cost. Although the current tight coupling between DRAM and PMM is likely a design choice for performance and convenience consideration, it may prohibit exploiting the full potential of power efficiency of PMM.

*Insight V: Energy-aware data placement would need to consider the high static power and the throttling effects from writes to PMM.*

*Insight VI: Non-temporal write in MemoryMode may result in bandwidth loss and high energy cost.*

### 4.4 Graph Applications

We further quantify the benefits of memory configurations on applications that require large memory capacity. We choose five popular graph applications from two well-known graph processing frameworks, i.e., Ligra [28] and GAP [1]. Each application uses several input problems whose memory footprint eventually scale beyond the DRAM capacity.

The algorithmic properties of these graph applications result in similar sensitivity to different memory configurations, even when different frameworks and implementations are used. Figure 9 presents the graph applications in the two frameworks using two input problems that have memory footprint smaller than the DRAM capacity. Thus, we can use the performance on DRAM configuration as the reference (the red dashed line) for normalizing the performance on the other three memory configurations. In this set of experiments, MemoryMode configuration shows similar performance as DRAM with little fluctuation for some kernels, indicating its effectiveness for graph applications with memory footprint fit in DRAM. PMM without DRAM caching, however, results in 2–18x slowdown depending on the application. The slowdown of an application is again consistent across the two frameworks. For instance, on both frameworks, triangle counting (TC) exhibit the lowest slowdown among all applications, i.e. up to 5x on Ligra and 2.5x on GAP framework. One reason for the low sensitivity is the relative high computation intensity in TC compared to other applications. In contrast, BFS exhibits high sensitivity when changing from DRAM to PMM in both frameworks, i.e. reaching up to 15x on Ligra and 18x on GAP framework. Finally, the DRAM-PMM interleave configuration highlights the importance of DRAM caching as its improvement compared to PMM configuration, about 2x speedup, is less impressive than Memory mode.

Large problems that exceed the DRAM capacity could still benefit from MemoryMode, but the improvement compared to PMM configurations diminishes as the problem size increases. Figure 10 presents the execution time of five graph applications in GAP framework using input problems that scale at a doubling rate from 35 to 270 GB. In BFS, CC and PR, the gap between DRAM and MemoryMode configurations continues increasing and shows a nonlinear increase at input s30, whose memory footprint exceeds DRAM capacity. As the problem size increases, the effectiveness of using DRAM as a cache to PMM continues decreasing as illustrated in Figure 11, where the execution time of PMM and DRAM-PMM-interleave configurations are normalized to that of MemoryMode configuration. At the largest problem, the performance gap decreased to 2x–6x while small inputs could have performance gap up to 18x. Note that DRAM-PMM-interleave configuration augments the total memory capacity by 192 GB compared to the MemoryMode configuration, which is substantial for the total memory capacity of a system. Thus, trading off a slight slowdown for enabling a large problem in DRAM-PMM-interleave configuration could be an feasible option for certain applications.

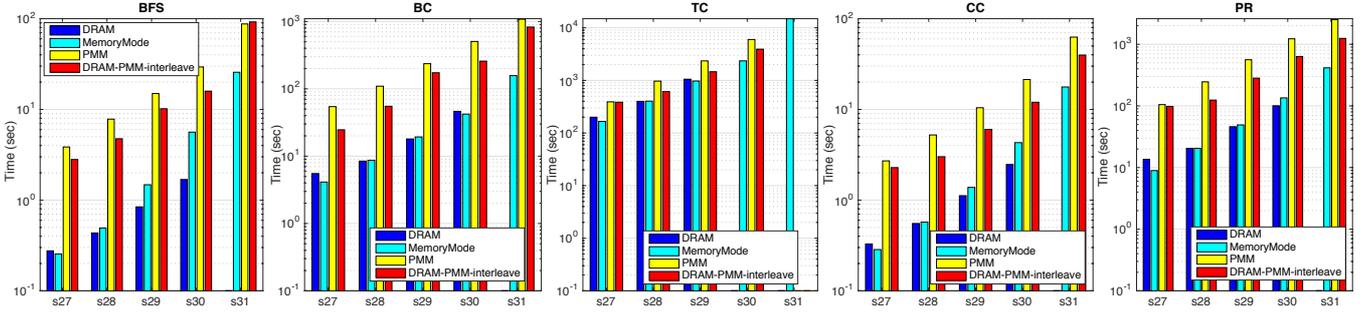

Figure 10: Execution time of five graph applications with increasing problem size (x-axis) in the GAP graph framework using four memory configurations. TC memory footprint exceeds DRAM capacity at s30.

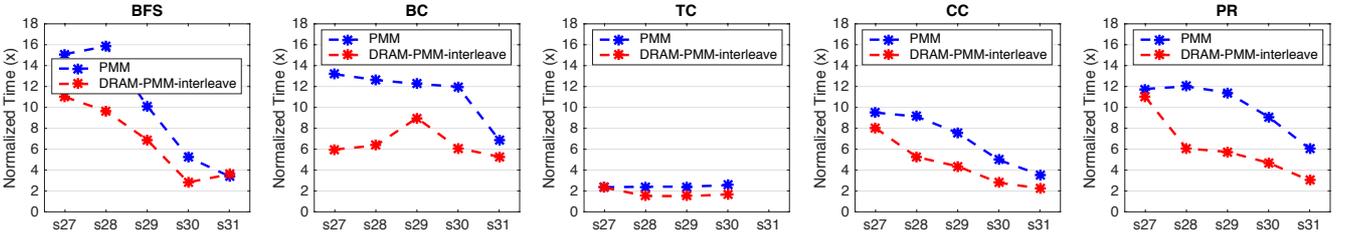

Figure 11: Performance gap between MemoryMode and two PMM configurations decreases at increased input problems.

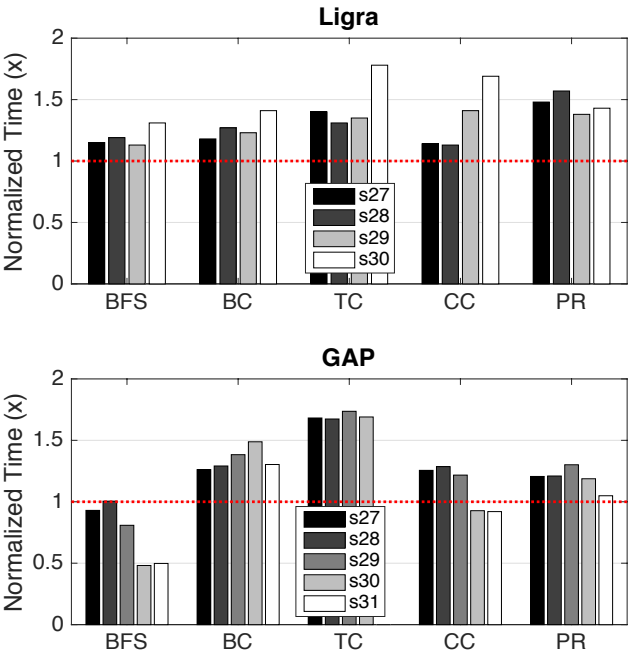

Figure 12: Normalized execution time using single socket w.r.t the execution time using two sockets.

NUMA effects are profound for graph applications without specific optimization to reduce remote-socket access. Section 4.1 and 4.2 have revealed the severe bottleneck in latency and bandwidth when accessing data on the remote socket. To quantify its realistic impact on applications, we compare single-socket and dual-socket executions of five applications in two frameworks. For single-socket execution, we use only cores and memories on one socket. The obtained execution time is then normalized to the execution using all cores and memories on both sockets. Figure 12 presents the relative performance of two execution modes, where results below the reference (the red dotted line) indicate that single-socket execution has shorter execution time than using two sockets.

Using cores and memories on two sockets does not always improve the performance. In both frameworks, less than 20% speedup is observed in dual-socket execution for small input problems compared to single-socket execution. Surprisingly, using two sockets could even slow down the execution of some applications, e.g., BFS and CC in the GAP framework. The slowdown by two sockets in these applications even increases when the problem size increases. In both frameworks, applications with low compute intensity like BFS, are more sensitive to the high overhead of remote-socket accesses. In contrast, applications that are more compute intensive like TC, can still benefit from the increased throughput on two sockets despite the NUMA penalties.

System Evaluation of the Intel Optane Byte-addressable NVM

Avoiding the severe performance loss due to the write throttling effects to PMM, and the high overhead of accessing remote socket becomes the priority in optimizing graph frameworks on similar DRAM-NVM memory systems. For instance, the Ligra framework performs an in-place sort on edges before computation. Graph edges are typically large data structures stored on PMM. Thus, the write-intensive sorting procedure is likely to be a bottleneck. Possible optimization could batch the sorting procedure in DRAM before placing data onto PMM to avoid frequent writes to PMM. Graph partition that maximizes local socket access and reduces remote accesses would also be feasible optimization technique [22].

*Insight VII: Graph partitioning among multiple sockets and write isolation from PMM would be critical and practical for performance.*

## 5 FINE-GRAINED MEMORY POLICIES

In this section, we employ two fine-grained memory allocation policies to improve the control of traffic distribution between NVM and DRAM. These policies require modifying applications. In return, they may bring more performance improvement than coarse-grained memory allocation. These policies could also workaround some performance bottlenecks in the memory configurations. We describe the allocation policies as follows.

**Bandwidth spilling** is a *DRAM-NVM-spilling Block Allocation* that returns a contiguous virtual memory space, which physically spills over two sockets and two memories (in numa configurations in Table 2). An allocation is divided into blocks, which are placed to sockets in a round-robin fashion. Each block spills from DRAM to NVM if the DRAM resource is exhausted. Thus, this allocation combines typical block allocation on NUMA machines to address the inter-socket bottleneck and also distributes traffic between DRAM and NVM to exploit the bandwidth.

**Write isolation** is an *NVM-aware-splitting Block Allocation* that returns a contiguous virtual memory space, which physically splits into multiple persistent structures over the two sockets (in PMM-fsdax configurations in Table 2). Blocks of one data structure are saved into multiple files and then spread over the Optane DC PMM on two sockets. Combined with thread affinity, this policy could mitigate inter-socket accesses and utilizes the aggregated throughput on two sockets.

### 5.1 Bandwidth Spilling

We have shown in Figure 4.1 and 5 that using DRAM as a cache to NVM could have high latency and low bandwidth when the data size approaches or exceeds DRAM capacity. To overcome the cache overhead and inter-socket delay, we explore fine-grained data placement control in App Direct mode. We derive a simple analytical model in Eq. 1 to guide the achievable bandwidth. In this model, $M_0$ represents the portion of the memory traffic to DRAM, and $BW_0$ and $BW_1$ represent the peak bandwidth to DRAM and PMM, respectively. Based on the model, we develop the bandwidth spilling block allocation routine.

$$BW_{tot} = \frac{1}{\frac{M_0}{BW_0} + \frac{1-M_0}{BW_1}} \quad (1)$$

Eq. 1 models the overall bandwidth as a nonlinear inverse variation function of the traffic distribution to PMM. Now assuming the total traffic is proportional to the data size and substitute $BW_0$ and $BW_1$ with the measured peak bandwidth of PMM and DRAM on two sockets, i.e., 78 GB/s and 204 GB/s respectively, we plot the theoretical bandwidth in the black dashed line in Figure 13. Note that we only consider read traffic for bandwidth-spilling because write accesses to PMM dramatically lower performance and should be isolated in DRAM as discussed in the next section.

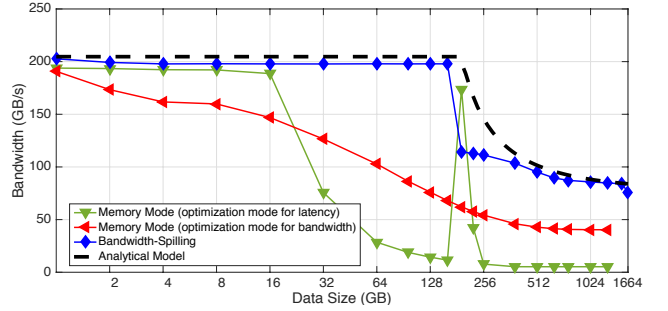

**Figure 13: Compare the bandwidth in App Direct mode using the bandwidth spilling policy with the optimization modes in Memory mode. Memory mode supports up to 1.28 TB data while the spilling policy reaches 1.5 TB.**

We evaluate the performance of the proposed policy using the accumulate benchmark and increase the data size to stretch the memory system. For small data size, Memory mode and our policy achieve similar bandwidth at about 200 GB/s. At about 32 GB, the two optimization options in Memory mode exhibit a reverse. The optimization mode for bandwidth starts outperforming the optimization mode for latency. For data larger than 256 GB, the option for latency in Memory mode decreases to 5 GB/s quickly while the optimization mode for bandwidth sustains at 40 GB/s. The bandwidth spilling policy achieves high bandwidth as predicted by the analytical model. When data size exceeds 1 TB, our policy still sustains 76 to 97 GB/s, about 2x improvement compared to the best performance in Memory mode. Additionally, our policy enables much larger data size at 1.54 TB, enabling 20% more data size than Memory mode.

### 5.2 Write Isolation

Previous sections have shown that write accesses to PMM result in severe performance degradation, high energy consumption, and write amplification. Separating write accesses to DRAM becomes critical, which is automatically achieved in Memory mode. In App Direct mode, one natural question is how much more improvement in performance and power is achievable if fine-grained policies are employed. To explore the potential for improvement, we employ NVM-aware-splitting allocation for read-intensive data structures and allocate write-intensive data onto DRAM in the STREAM [20] benchmark. We perform experiments on four dual-socket configurations in Table 2.

Our results show that this write isolation policy improves memory bandwidth at large data size and avoids the throttling effect due to writes to PMM. At medium data sizes in Figure 14, MemoryMode configuration can effectively bridge the performance gap between DRAM and PMM, achieving 46–89% DRAM bandwidth. The write isolation policy starts outperforming the MemoryMode when data size is larger than 32 GB. At the largest data size, it sustains 83 GB/s bandwidth while the Memory mode reaches 27 GB/s. Note that in Figure 5, using a read-only benchmark, MemoryMode achieves much higher utilization of DRAM bandwidth than using the stream copy benchmark. We attribute the difference to the *throttling effect*, where evicting dirty cache lines in DRAM results in long latency writes to PMM, indirectly impacting the read access to that cache line. The write isolation policy could effectively bypass this throttling effect to improve performance. The trade-off between the porting efforts in fine-grained policies and the performance in Memory mode depends on the data size and the complexity in managing data explicitly in an application.

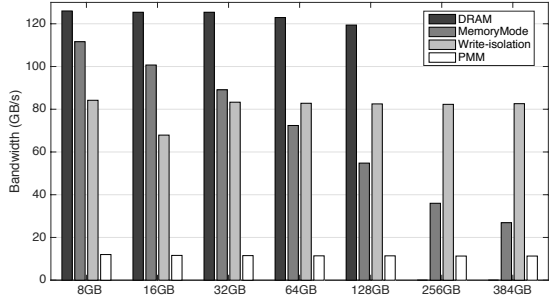

Figure 14: Compare the bandwidth in App Direct mode using write isolation policy with DRAM, MemoryMode, and PMM configurations.

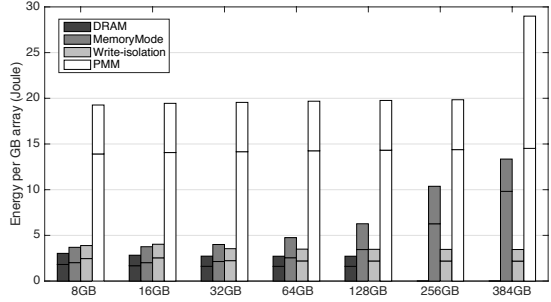

Figure 15: The total energy on two sockets for each gigabyte of data in the stream benchmark. Each bar is partitioned into CPU (bottom) and memory energy (top).

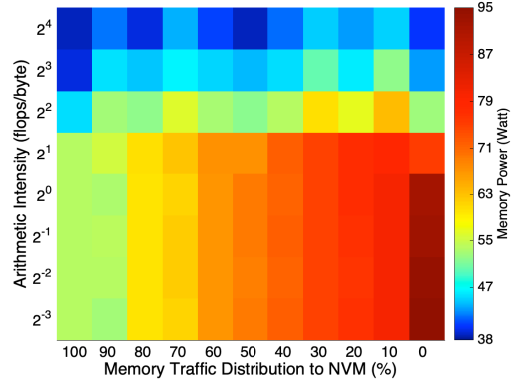

Figure 16: Measured memory power on one socket at different traffic distribution between NVM and DRAM (x-axis) and arithmetic intensities (y-axis).

The write-isolation policy reduces energy cost by up to 8.4 times compared to PMM and 3.9 times compared to MemoryMode. In Figure 15, MemoryMode shows increasing energy cost per gigabyte of data when the total data size increases. The CPU energy constitutes only 55% of the total energy at small data size but increases to 74% at the largest data size. As the bandwidth of MemoryMode is decreasing when the data size increases, i.e., the dynamic CPU power should be decreasing, we attribute the increased CPU energy mostly due to the increased static energy for prolonged execution time. For energy-aware applications, the potential energy saving from the write-isolation policy could well justify the porting efforts.

### 5.3 Traffic Distribution

In this section, we sweep the arithmetic intensities in workloads to explore the real impact of an NVM-DRAM memory system on different workloads. Arithmetic intensity is defined as the number of (floating-point) operations per byte from the memory subsystem [33]. In general, high arithmetic intensity results in low sensitivity to the memory system and vice versa. We employ a modified stream accumulate benchmark to sweep the arithmetic intensity and to establish the *roofline model* [33] of theoretical peak performance on our platform. Note that this exploration focuses on peak performance, and thus studies read traffic only because writes to NVM severely reduce performance. Our objective of this study is to control the memory traffic to DRAM and NVM at fine grain to understand how to adapt the traffic distribution based on the application sensitivity, and eventually achieve better performance or energy efficiency. Hence, we combine the roofline model and power consumption to established the power-line, and arch-line model [4] for guiding the search for the optimal distribution.

The first part of our exploration is to establish power consumption at different arithmetic intensities and traffic distribution. Figure 16 presents a heat map of memory power on one socket, including static and dynamic power. In general, memory power in all distributions decreases steadily when arithmetic intensity increases along the y-axis. Memory-intensive workloads, whose intensity is lower than $2^1$ on the y-axis, have power consumption directly increase along the x-axis, i.e., increased traffic distribution to DRAM. For the most memory-intensive workload ($2^{-3}$ on the y-axis), distributing all memory traffic to DRAM (the right end on the x-axis) results in the highest power consumption, at about 95 watt. With 100% distribution to NVM, the memory consumption is 54 watt. Note that this 54 watt power still includes the static power from DRAM DIMMs. Applications with medium and high arithmetic intensity ($2^1$–$2^4$ on the y-axis), however, may consume more power when the traffic distribution is skewed. For these workloads, adapting the memory traffic distribution to PMM could lower power consumption.

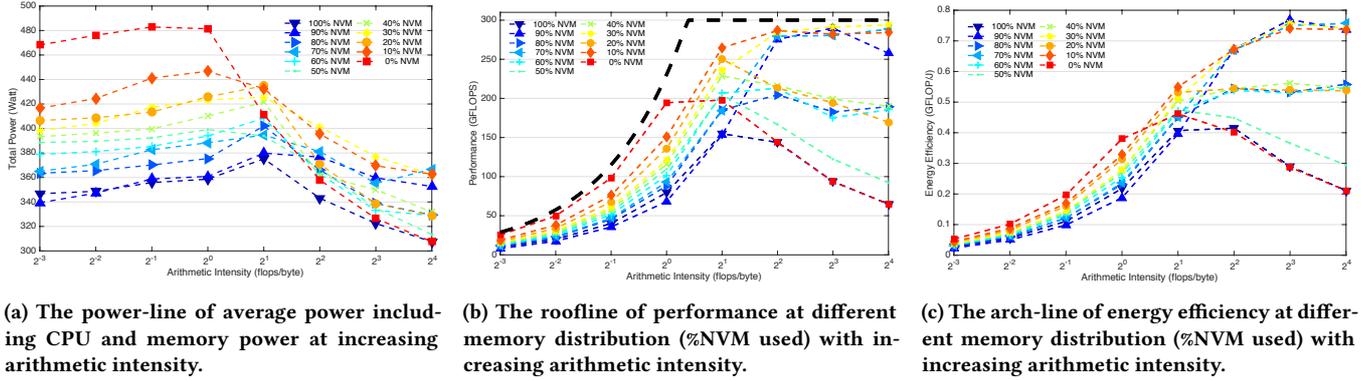

(a) The power-line of average power including CPU and memory power at increasing arithmetic intensity.

(b) The roofline of performance at different memory distribution (%NVM used) with increasing arithmetic intensity.

(c) The arch-line of energy efficiency at different memory distribution (%NVM used) with increasing arithmetic intensity.

Figure 17: Established models for peak performance and power efficiency at various arithmetic intensities.

The power-line [4] usually depicts a power peak at a certain arithmetic intensity when the arithmetic intensity increases from low to high. Figure 17a reports the total power consumption of the platform, including CPU and memory on both sockets. We can observe the power peak at $2^1$ on the x-axis for most memory distributions except 0% and 10% distribution. A 0% distribution indicates that all traffic goes to DRAM and we find that it consumes over 480 watt power. Unlike other distributions, this distribution shows no power peak, which is possibly due to the power capping on the platform. Throughout different arithmetic intensities, the gap between 0% and 100% distribution could even reach 125 watt. Interesting, at low compute intensity, distributing as low as 10% traffic to NVM saves up to 40 watt power, resulting in a wide gap between the 0% and 10% lines on Figure 17a.

The roofline model [33] links the memory bandwidth and operation intensity to theoretical peak performance for exploring optimization opportunities. Figure 17b presents the derived roofline for our platform. The model indicates that the limiting factor of performance changes from the memory system to the computing capability at $2^0$ to $2^1$ arithmetic intensity. Below this, full distribution to DRAM brings the highest performance. Once the arithmetic intensity is higher than $2^1$, a full distribution to either DRAM or NVM causes suboptimal performance compared to other distributions. Although high arithmetic intensity is expected to result in low sensitivity to the memory system, our results show that the traffic distribution between NVM and DRAM could still impact the performance. Finally, we derive the arch-line of energy efficiency [4] in Figure 17c to study the impact of traffic distribution on energy efficiency. The results again diverge at arithmetic intensity $2^1$, where distributing 10% or 90% traffic to NVM brings higher efficiency than other distributions in traffic.

## 6 RELATED WORK

Extensive works have proposed different materials and architectures for implementing non-volatile memories, including spin torque transfer RAM (STT-RAM), resistive RAM (RRAM), and phase changing memory (PCM) [8, 17, 26, 30]. While these works demonstrate prototype designs, the Optane DC PMM in this study is the first commercially available hardware that provides enormous memory capacity.

Many studies have extensively investigated software techniques for improving application performance on heterogeneous memory systems even before the NVM hardware is available [5, 23, 34]. These works identify data structures or pages that are critical for performance and manage data placement between different memories, either statically or at runtime. Another group of studies focuses on identifying future system designs for improving application performance or energy consumption [14, 16, 24]. As the hardware was unavailable, most works used software or hardware emulators or cycle-accurate simulators for evaluation.

Since the Optane DC PMM becomes available, several groups have performed extensive studies from different perspectives. [13] uses representative in-memory database workloads, which are critical for data centers. Their work also shows the advantage of NVM-specific file system [35]. [6] optimizes the Galois framework [21] to mitigate the NUMA effect in memory mode. They also compared the scalability of Galois on a single machine with the distributed-system implementation. In addition to their findings, our work provides an evaluation of power and energy efficiency at various memory configurations as well as fine-grained traffic controls between NVM and DRAM.

## 7 CONCLUSION

Byte-addressable NVMs are a promising new tier in the memory hierarchy on future large-scale systems. In this work, we evaluated the first commercially available byte-addressable NVM based on the Intel Optane® DC™ technology. We expect that memory configurations that require no application modifications would likely be the first deployment efforts. Thus, our evaluation quantified the performance of eight memory configurations, and more importantly, provide guidelines for selecting suitable configurations for applications. Our evaluation of five graph applications shows that DRAM-cached NVM could bring reasonable performance for large graphs. The second part of our study explores the potential of further improvement with fine-grained control of the memory traffic between NVM and DRAM. Our results show that Optane is advantageous in enabling power-efficient workloads when data is carefully partitioned and placed on different memories. With porting efforts to support bandwidth-spilling and write-isolation policies, applications could achieve higher bandwidth and lower energy cost than the coarse-grained memory configurations. Finally, our work provides first-hand insights for optimizing applications on the emerging memory systems that feature byte-addressable NVM.

## 8 ACKNOWLEDGMENTS

This work was performed under the auspices of the U.S. Department of Energy by Lawrence Livermore National Laboratory under contract No. DE-AC52-07NA27344 and was supported by the DOE ECP project. This document was prepared as an account of work sponsored by an agency of the United States government. Neither the United States government nor Lawrence Livermore National Security, LLC, nor any of their employees makes any warranty, expressed or implied, or assumes any legal liability or responsibility for the accuracy, completeness, or usefulness of any information, apparatus, product, or process disclosed, or represents that its use would not infringe privately owned rights. Reference herein to any specific commercial product, process, or service by trade name, trademark, manufacturer, or otherwise does not necessarily constitute or imply its endorsement, recommendation, or favoring by the United States government or Lawrence Livermore National Security, LLC. The views and opinions of authors expressed herein do not necessarily state or reflect those of the United States government or Lawrence Livermore National Security, LLC, and shall not be used for advertising or product endorsement purposes. LLNL release LLNL-PROC-777377.